\documentclass[preprint,showpacs,preprintnumbers,amsmath,amssymb,aps,prb,floatfix]{revtex4-1}

\usepackage{graphicx}
\usepackage{dcolumn}
\usepackage{bm}

\input colordvi

\begin{document}



\title{Density Functional Theory Study of
the Entangled Crystal, Magnetic, and Electronic Structures of PuGa$_3$}

\author{Sven P. Rudin}
\affiliation{Los Alamos National Laboratory, Los Alamos, New Mexico 87545, USA}

\date{\today}

\begin{abstract}
Systematically studying the crystal, magnetic, and electronic structures
of PuGa$_3$ with density functional theory (DFT)
reveals the entanglement of the three types of structure.
Magnetic structure affects the energy more strongly than crystal structure.
For DFT to correctly order the crystal structures in agreement with experiment
requires special treatment of the electronic correlation in the $5f$ states,
exemplified here by the GGA+U approach.
The upper and lower Hubbard bands change with increasing $U$
in very dissimilar ways for the two most different crystal structures.
The results suggest the effectiveness of using magnetic structure to simulate correlation
effects in the actinides depends on both the magnetic and the crystal structure.
\end{abstract}

\maketitle

\section{introduction}

PuGa$_3$ appears in two crystal structures
with similar electronic structures
but different magnetic structures.
Ellinger {\it et al.} noted in 1964 the appearance of the two crystal structures
and identified the low-temperature (low-$T$) form as hexagonal and
isostructural with Ni$_3$Sn;\cite{ellinger64}
the following year Larson {\it et al.} identified the high-temperature (high-$T$) form
as a 12-layer rhombohedral close-packed structure
with space group R$\bar{3}$m.\cite{larson1965}
Four decades later,
magnetic measurements revealed
the low-$T$ form to be an antiferromagnet ($T_N=24$ K)
and
the high-$T$ form to be a ferromagnet ($T_C=20$ K),
and specific heat measurements suggest for both phases
an electronic structure with heavy fermion character.\cite{boulet05}

These characteristics place PuGa$_3$ in compelling relation
to other heavy fermion systems of significant interest.
The electronic specific heat coefficient $\gamma$,
a measure of the electronic density of states at the Fermi energy,
has values
(220 and 100 mJ/mol K$^2$ for low-$T$ and high-$T$, respectively)
similar to the heavy fermion superconductors
PuCoGa$_5$ and PuRhGa$_5$.\cite{javorsky2007}
Neither of these are magnetic experimentally,
though electronic structure calculations favor antiferromagnetic
order.\cite{opahle2004}
With rather delocalized $5f$ electrons, PuGa$_3$ lies between
PuCoGa$_5$ and $\delta$-Pu,\cite{boulet05}
which has more localized $5f$ electrons but
also shows no localized magnetic moments.\cite{lashley05}

The combination of magnetic structure and heavy fermion behavior
in PuGa$_3$ suggest a challenging system for electronic structure
calculations.
The electronic structure of Pu, many Pu compounds, and some other actinide systems
requires special attention to be paid to the strong $5f$ electron correlation.
Calculations with ``standard'' density functional theory (DFT) methods,
which involve limited approximate treatments of the electronic correlation,
favor an antiferromagnetic structure,\cite{postinov00, wang00, soderlind01}
and some aspects are even better modeled with disordered local moments
or approximations thereof.\cite{niklasson03, soderlind04a}
Experimentally, pure Pu shows no signs of magnetic moments.\cite{lashley05}
The breaking of spin symmetry in DFT calculations delivers a static approximation of the
spatial separation experienced by dynamically correlated electrons.
As a result, calculations allowing a localized magnetic moment can be
used to explore nonmagnetic
aspects of Pu and Pu compounds without introducing material-dependent
parameters.
The existence of magnetic structures in PuGa$_3$ entangles the magnetic
moments and the electronic correlation, which,
along with their entanglement with the observed crystal structures,
motivates this study.

The crystal structures of PuGa$_3$
can be viewed as close-packed PuGa$_3$ planes
with different stacking sequences.\cite{larson1965}
The low-$T$ structure follows an AB sequence,
as does hexagonal close packed (hcp);
stacking in the high-$T$ structure progresses as ABABCACABCBC
(with some in-plane distortions away from the perfect close-packed planar structures).
This layered, close-packed nature already appears in crystal structures of pure Pu:
the face-centered cubic structure of $\delta$-Pu exhibits ABC stacking,
the Crocker pseudostructure for $\alpha$-Pu follows from the $\alpha$ structure's
repeating two planes of a distorted hexagonal structure,\cite{crocker1971}
and the orthorhombic structure of $\gamma$-Pu exhibits close-packed Pu planes
stacked such that Pu atoms in one plane sit above bonds in the plane underneath
(giving rise to an ABCD stacking pattern).
The close-packed PuGa$_3$ planes correspond to the close-packed Pu planes
with ordered substitutional placing of Ga.

These stacking sequences of close-packed planes
(excluding that of $\gamma$-Pu)
can also be written as sequences of shifts
between planes:
AB, BC, CA being shifts to the right (R) and
AC, CB, BA being shifts to the left (L).
ABC stacking always shifts in the same direction (RRRR),
AB stacking alternates between the two directions (RLRL), and
ABABCACABCBC stacking, rewritten as (ABCA)(CABC)(BCAB),
reverses direction once every four planes (RRRL).
The missing unique pattern with four shifts,
RRLL, is ABCB stacking, which corresponds to double hcp (dhcp),
exemplified by $\alpha$-La.

While the R and L shifts are equivalent,
a stark difference exists between a plane that
links two shifts with the same direction and one
that sits at a reversal in the direction.
The local environment of the atomic sites in the ideal
close-packed lattices has twelve nearest neighbors
in both cases. 
Sites in a plane 
between two identical shifts have
the inversion symmetry, while
those between two opposite shifts do not.
The lack of inversion symmetry disrupts an otherwise
straight line of bonding oriented 60$^\circ$ to the planes.
A natural order of the four crystal structures arises:
ABC stacking has no disruptions,
AB stacking has disruptions in every plane,
and the remaining two stacking sequences lie in between.

The work presented here applies DFT to
reveal the interplay between crystal structures based on these
four structural patterns,
a series of magnetic structures, and the resulting electronic structures.
Starting from ``standard'' DFT in the generalized gradient approximation (GGA),
calculations furthermore explore
the effects of adding either spin-orbit coupling or a
Hubbard $U$ (in the GGA+U method).
The calculations presented here set aside
thermal effects, in particular those due to phonons.
Preliminary calculations of the phonons and their contribution to the
free energy suggest they cannot make the low-$T$ phase more favorable
in the GGA to DFT without specifically addressing $f$ electron correlation.

\section{method}

The DFT calculations employ
the {\sc VASP} package.\cite{kresse96, kresse99}
They make use of
the generalized gradient approximation (GGA) of
Perdew, Burke, and Ernzerhof.\cite{PBE96}
The Pu($5f, 6d, 7s$) and Ga($4s, 4p$) electrons are treated in the valence
using a plane-wave basis and
with projector-augmented wave potentials.\cite{blochl94a}
The calculations employ Methfessel-Paxton smearing (with width 0.1 eV),
a k-point mesh of density 40~\AA$^{-1}$,
and an energy cutoff of  400 eV.
The self-consistent cycles are converged to within 10$^{-5}$ eV.
Calculations aimed at improving the treatment of the
the on-site Coulomb repulsion between 5$f$ electrons
use an effective Hubbard parameter $U$
in the rotationally invariant form of Dudarev {\it et al.}.\cite{dudarev98}
In this form the Hubbard parameter $U$ and the exchange parameter $J$ 
appear only in the difference $U-J$,
throughout this report the difference is referred to as $U$.
Calculations that include the effects of spin-orbit coupling do so
in the noncollinear mode of {\sc VASP},\cite{hobbs00,marsman02}
the implementation follows the approach of
Kleinman and MacDonald, Picket, and Koelling.\cite{PhysRevB.21.2630,macdonald80}

The calculations optimize crystal structures that
start as ideal close-packed planes with one Pu and three Ga atoms,
stacked according to one of the four patterns described above.
Relaxation of the structures retains the overall layered structure,
but displacements within the planes make initially equivalent planes
lose their exact equality.
The size of the unit cell, in particular the number of planes (between two and twelve),
follows from the particular pattern and the magnetic structure
used to seed the calculations.
The latter either has all spins in the same direction for
the ferromagnetic (FM) structure,
or spins that switch direction every one, two, three, four, or six planes.
These arrangements define spin wave structures with wave vectors $\mathbf{q}$
of magnitude $\frac{1}{1}$, $\frac{1}{2}$, $\frac{1}{3}$,
$\frac{1}{4}$, $\frac{1}{6}$, and, in the FM case, $\frac{1}{\infty}$,
scaled by $\frac{\pi}{c_0}$, where $c_0$ represents the interplanar
spacing.
Additional magnetic structure within the close-packed plane affects
the results, but these are not reported here, other than to note that
their energy lies above that of the antiferromagnetic
(AFM, $|\mathbf{q}| = 1\frac{\pi}{c_0}$) state.

\section{results}

The results from three approaches appear in the following
three subsections.
Sections \ref{GGA} and \ref{GGAplusSOC} report the results of
DFT calculations in the GGA without and with spin-orbit coupling,
respectively, for the four crystal structures in a sequence of magnetic
states.
Section \ref{GGAplusU} focuses on results of the GGA+U method
applied to the low-$T$ and cubic structures in the AFM state.
Table \ref{tab:opt} summarizes the energies, volumes and $c/a$ ratios
calculated in the three approaches for the four crystal
structures in the AFM state.

\subsection{\label{GGA} DFT in the GGA}

\begin{figure} 
\includegraphics[width=8.5cm]{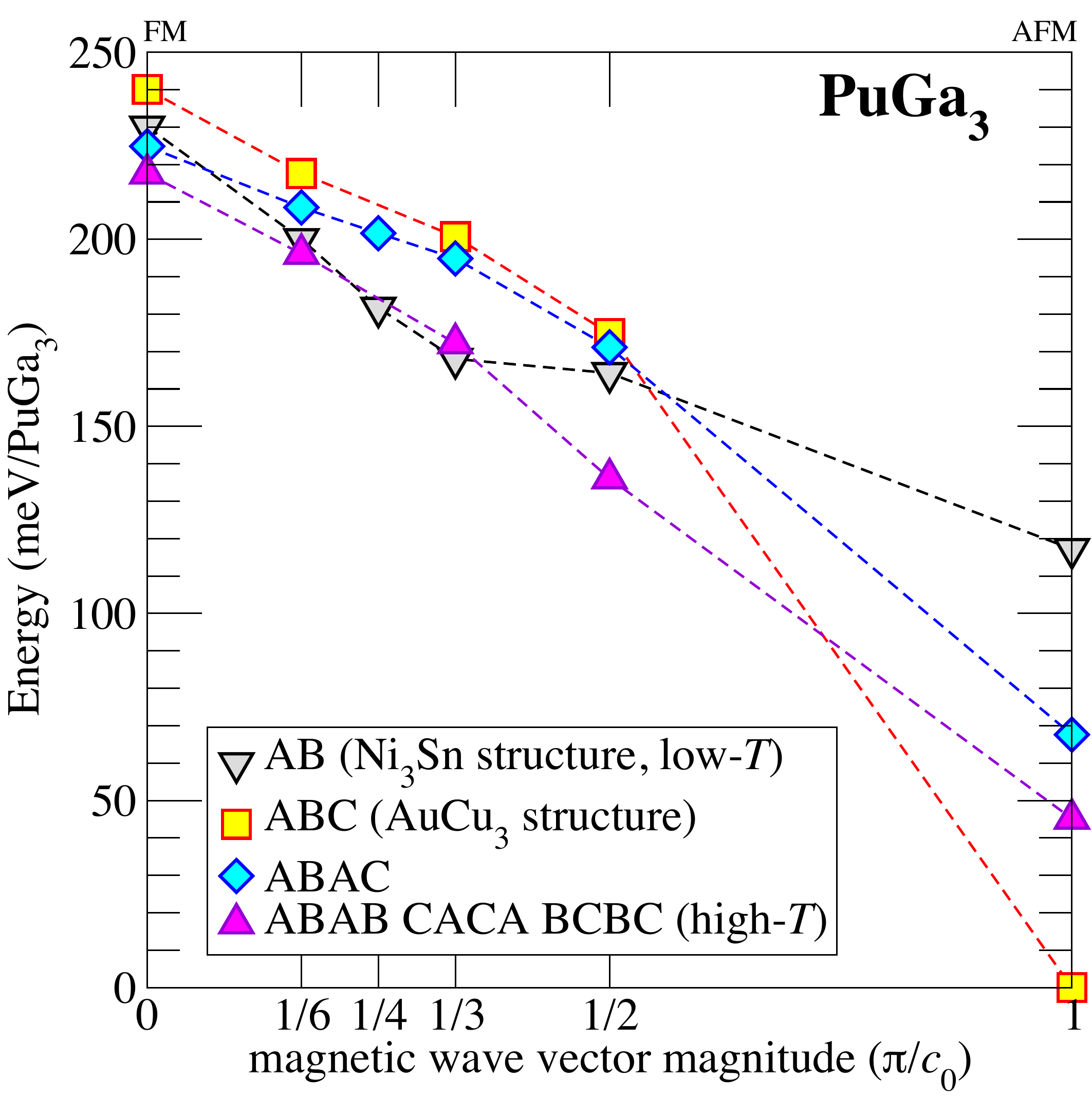}
\caption{\label{fig:ABCafvary}
(Color online)
Calculated dependence of energy on stacking and magnetic structure for PuGa$_3$
using GGA ($U=0$).
Stacking denotes initial crystal structure; upon relaxation the planes with the same
letter are no longer necessarily equivalent.
Stacking direction corresponds to body diagonal of the conventional AuCu$_3$
crystal structure unit cell; the lowest energy appears for G-type antiferromagnetism
(AFM(G)).
Dashed lines serve to guide the eye.
}
\end{figure}

Figure \ref{fig:ABCafvary} shows the interplay between the four
crystal structures and the magnetic structures using the GGA to DFT.
All four stacking sequence patterns show a preference for the
magnetic structure with the shortest spin wave length, the AFM state.
The ordering of energies of the crystal structures in the AFM state
correlates with the order arising from the number of
changes in R and L shifts mentioned in the introduction.
With this magnetic structure, the experimentally observed low-$T$ phase
lies highest, 117 meV/PuGa$_3$ above the favored structure
with ABC stacking.
This cubic structure is observed for PuIn$_3$ and is
often considered a building block for the layered superconductors
PuCoGa$_5$, CePt$_2$In$_7$, Ce$_2$RhIn$_8$, etc.
The preference for this cubic structure appears only for the
magnetic structure with the shortest spin wave length;
for longer spin wave lengths it lies higher than the other states
(albeit by small amounts).
Among the FM states the high-$T$ phase lies lowest,
the slight 13 meV/PuGa$_3$ difference to the low-$T$ phase
suggests the importance of thermal effects.

The optimized structures agree reasonably well with experimental
volumes, while the optimized $c/a$ ratios consistently lie above the
experimental values.
The AFM volume calculated for the low-$T$ structure is only 1\% smaller,
but the $c/a$ ratio is close to 6\% larger than the experimental
value (see Table \ref{tab:opt}).
The AFM volume calculated for the high-$T$ structure is 0.25\%
smaller than the experimental value, and the $c/a$ ratio is
close to 4\% larger than the experimental value.
The FM volume calculated for the high-$T$ structure is 1\% larger,
and the $c/a$ ratio is 3\% larger than the experimental value.
The distances from Pu to nearest Ga atoms (located in adjacent planes)
differ by negligible
amounts between the calculated and experimental high-$T$ structures.
The larger calculated $c/a$ ratio does affect the angle spanned by
a Pu atom and two Ga atoms in adjacent planes, decreasing it by as much as 13\%.

\begin{figure} 
\includegraphics[width=8.5cm]{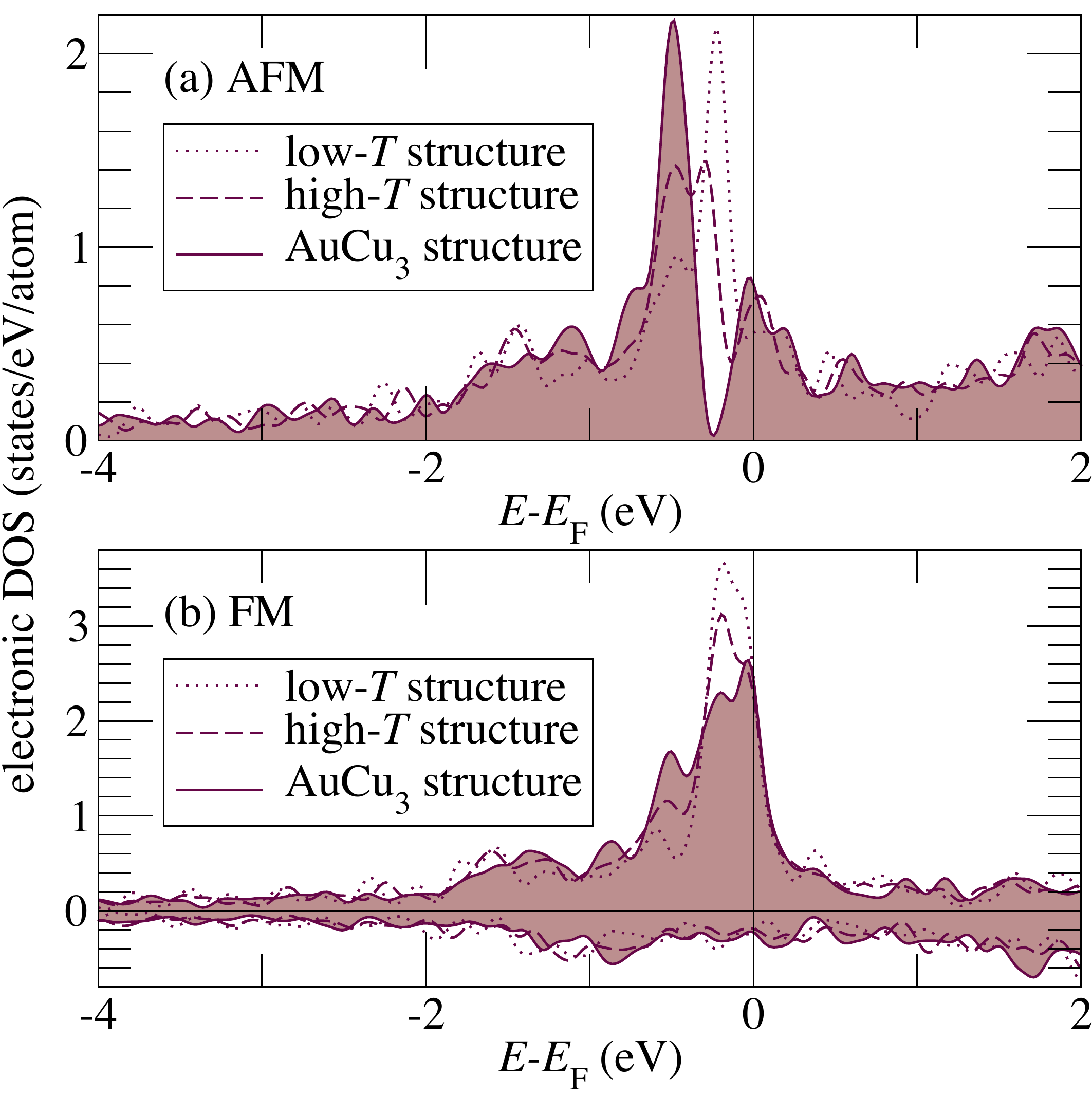}
\caption{\label{fig:GGAeDOS1}
(Color online)
Calculated electronic densities of states (DOS) near the Fermi energy $E_\text{F}$
using the GGA to DFT  ($U=0$) for
low-$T$, high-$T$, and AuCu$_3$ crystal structures with (a) AFM and (b) FM structure.
Only the DOS for one spin orientation appears for AFM.
}
\end{figure}

Figure \ref{fig:GGAeDOS1}(a) compares calculated total electronic
densities of states (DOS) and
suggests why the cubic structure
appears more favorable in the AFM state.
The low-$T$, high-$T$, and cubic structure differ significantly in
the highest occupied states.
The low-$T$ and cubic structure both exhibit a single peak,
but the cubic structure has it almost 0.3 eV further below
the Fermi level $E_\text{F}$.
The high-$T$ structure exhibits a double peak centered between the other two
structures.
While the band energy is only one part of the total energy,
this ability of the structures to push states down and away from
$E_\text{F}$ corresponds to their order in total energy.

Figure \ref{fig:GGAeDOS1}(b) plots the analogous comparison
for the three structures in the FM state.
Compared to the AFM state, the peaks
appear much more similar for the three structures than in the AFM state.
Accordingly, the total energies for the FM state
differ by smaller amounts compared to the AFM state.
The peaks sit closer to $E_\text{F}$ in the FM state, concurring with
the energies of the FM state lying higher than those of the AFM state.

\begin{figure} 
\includegraphics[width=8.5cm]{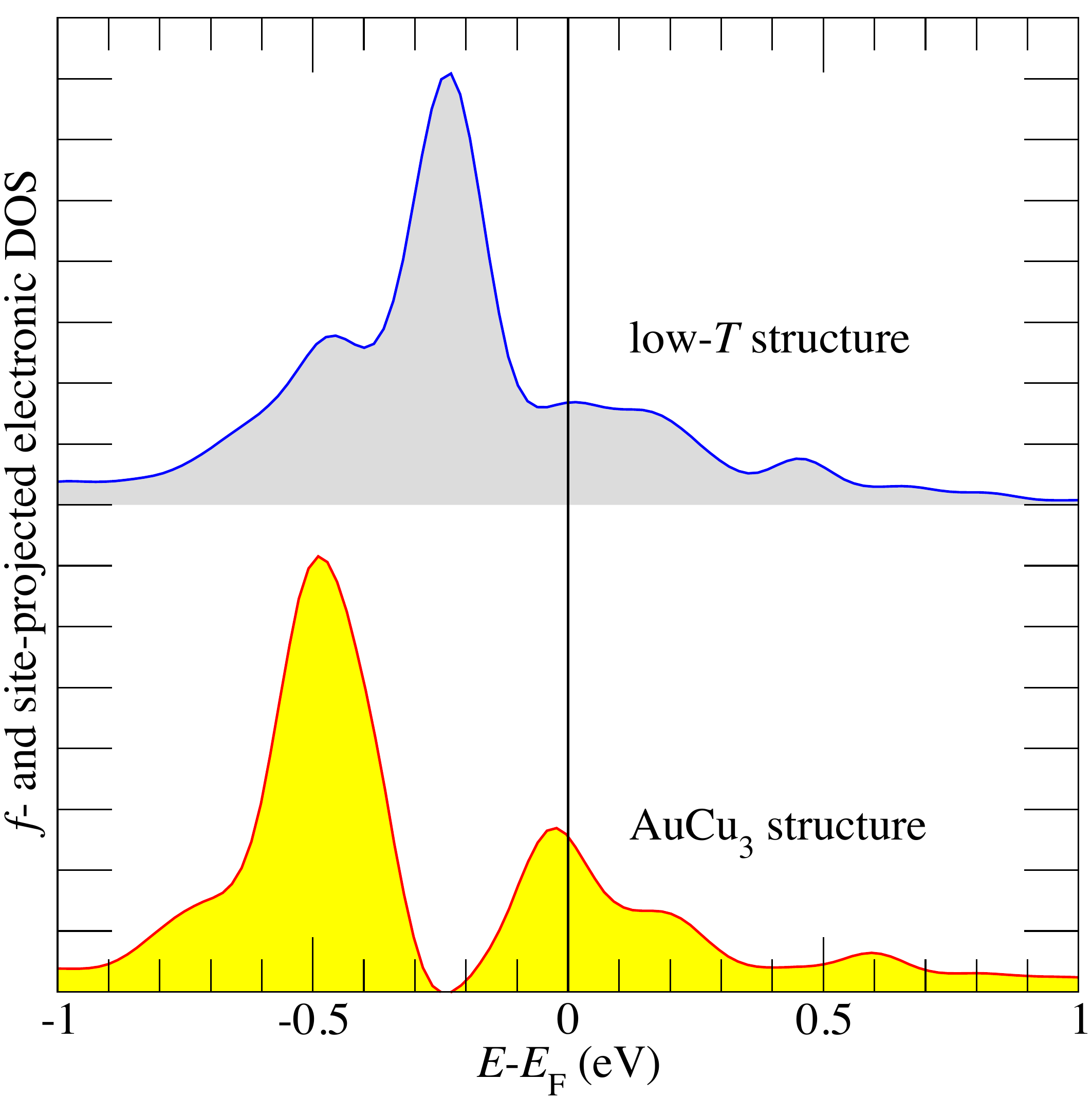}
\caption{\label{fig:fDOS1}
(Color online)
Calculated electronic DOS near $E_\text{F}$ 
projected on a Pu site with $f$ character
using the GGA to DFT ($U=0$) for
low-$T$ and AuCu$_3$ crystal structures with AFM magnetic structure.
The plotted DOS represent the majority spin on the Pu site.
}
\end{figure}

Figure \ref{fig:fDOS1} shows the $f$ symmetry character 
(projected out on a Pu site)
of the electronic DOS calculated for the low-$T$ and AuCu$_3$
crystal structures with AFM magnetic structure.
The $f$-projected peaks correspond to the peaks in Fig.\ \ref{fig:GGAeDOS1}(a).
The projected DOS are identical for all sites in each case, as expected
given the sites' identical environments:
each site has the same structural environment
and nearest neighbors with opposite spin.
The structural environment differs between the two cases,
the AuCu$_3$ crystal structure's inversion symmetry allows the
$f$-projected peaks to be pushed down lower.
The less symmetric local environment in the low-$T$ structure
makes it less atomic-like, requiring the $f$ electrons to
hybridize more.

\begin{figure} 
\includegraphics[width=8.5cm]{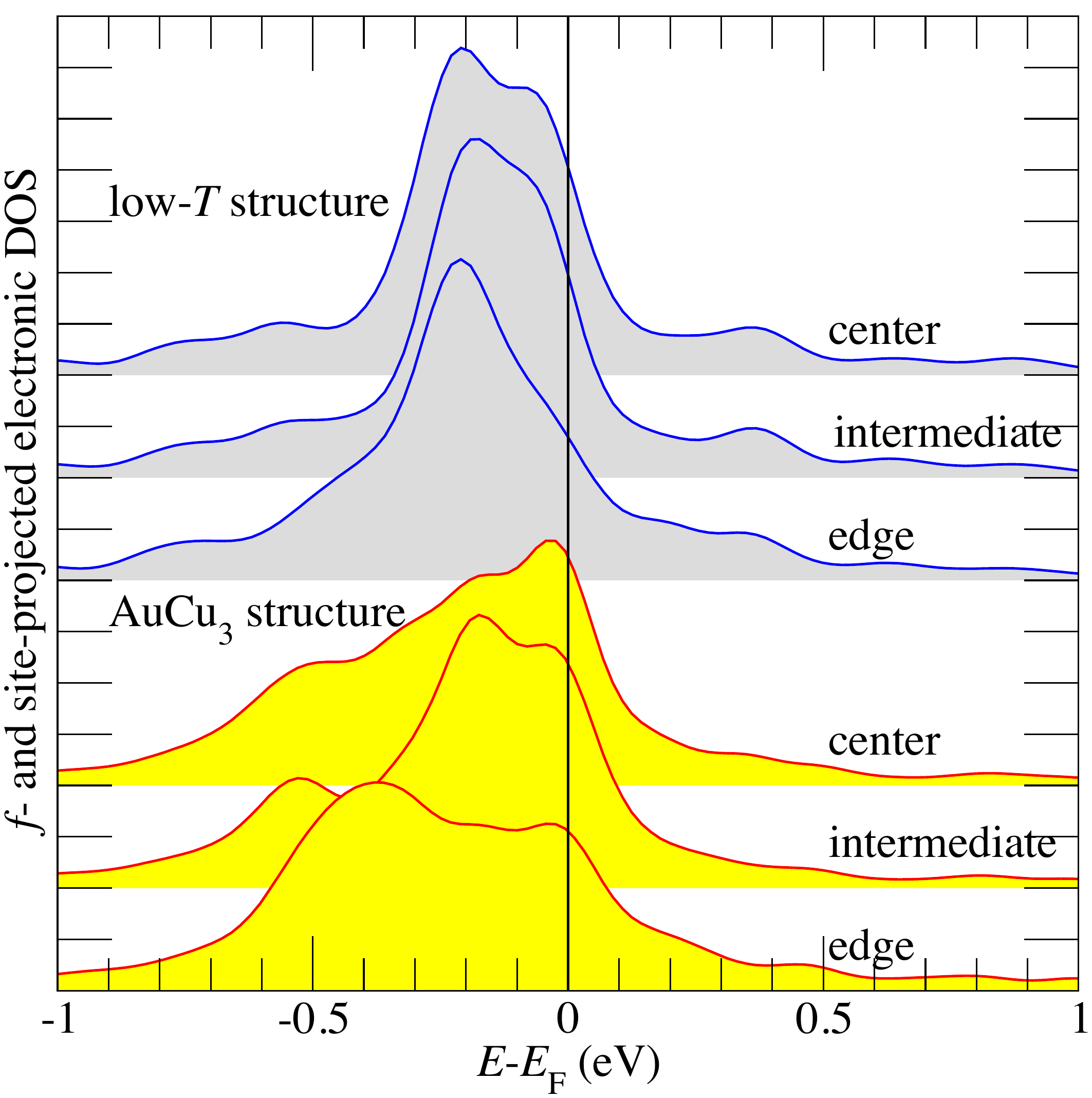}
\caption{\label{fig:fDOS2}
(Color online)
Calculated electronic DOS near $E_\text{F}$ 
projected on Pu sites with $f$ character
using GGA ($U=0$) for
low-$T$ and AuCu$_3$ crystal structures with magnetic structure
that has spin wave length spanning twelve planes.
The plotted DOS represent the majority spin on each site.
In terms of geometry, all sites are equivalent for each crystal structure.
They differ depending on where they sit within the magnetic structure:
adjacent to the spin flip (``edge''),
one layer farther in (``intermediate''), or
most distant to the spin flip (``center'').
}
\end{figure}

Figure \ref{fig:fDOS2} compares the $f$ symmetry character
projected out on Pu sites from the
electronic DOS calculated for the low-$T$ and AuCu$_3$ crystal structures
in the magnetic state with spin wave vector magnitude $\frac{1}{6}\frac{\pi}{c_0}$.
This choice of spin wave vector stems from the differences it
reveals among the Pu sites,
unlike the ferromagnetic structure where all sites (within each crystal structure)
remain equivalent.
For the low-$T$ structure the projected electronic DOS differs only slightly between
the three types of sites,
a slight shift down from $E_\text{F}$ occurs closer to the edge of the
magnetic subdivision.
The cubic structure shows dramatic differences between the three types
of sites:
all show a projected electronic DOS hugging $E_\text{F}$ from below,
and only the site at the edge of the magnetic subdivision appears able
to spread a significant amount down several tenths of an eV.
In the FM state, the $f$-projected DOS on any of the sites closely resembles
the $f$-projected DOS shown here for center atoms.

\subsection{\label{GGAplusSOC}
including spin-orbit coupling}

\begin{figure} 
\includegraphics[width=8.5cm]{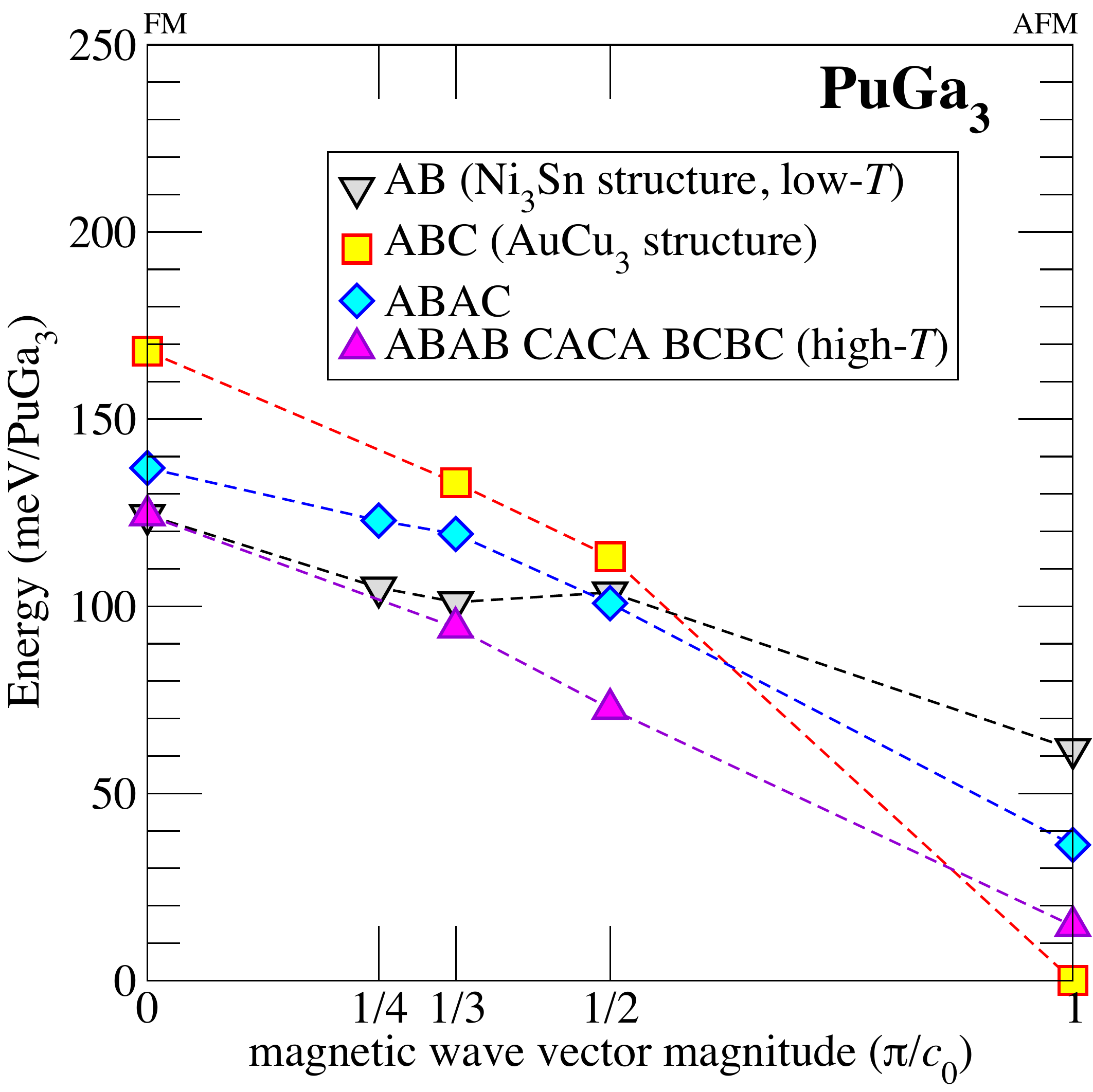}
\caption{\label{fig:ABCafvarySOC}
(Color online)
Calculated dependence of energy on stacking and magnetic structure for PuGa$_3$
using GGA ($U=0$) and including spin-orbit coupling.
Dashed lines serve to guide the eye.
Notation follows Fig.\ \ref{fig:ABCafvary}.
}
\end{figure}

Figure \ref{fig:ABCafvarySOC} shows the interplay between the four
crystal structures and magnetic structures using the GGA to DFT
and including spin-orbit coupling.
The inclusion of spin-orbit coupling reduces the energy differences overall,
hence Figure \ref{fig:ABCafvarySOC} appears
much like a scaled version of Fig.\ \ref{fig:ABCafvary}.
The AuCu$_3$ crystal structure in the AFM state remains the most
favored, in the FM state it remains the least favored.

Results from calculations that include spin-orbit coupling repeat
the correlation between which structure is energetically favored
and its ability to push electronic states down and away from $E_\text{F}$.
With spin-orbit coupling,
the electronic DOS of the low-$T$ and AuCu$_3$ crystal structure differ
from one another less than in Fig.\ \ref{fig:GGAeDOS1}, but the more
favored AuCu$_3$ crystal structure still succeeds better at pushing
electronic states to lower energies.

\subsection{\label{GGAplusU}
including a Hubbard U}

Table \ref{tab:opt} shows how treating the on-site Coulomb repulsion
between $5f$ electrons with a Hubbard U changes the
ranking of crystal structures.
Setting $U=3$ eV reverses the sequence in energy 
from the GGA result (with or without spin-orbit coupling):
the low-$T$ crystal structure becomes most favored while the
AuCu$_3$ crystal structure becomes the least favored.
The high-$T$ and ``$\alpha$-La'' structures remain in between and
switch their order as well.
Comparison of the energies for the different crystal structures only
has meaning for each value of $U$ individually, which is somewhat
unsatisfactory since the different crystal structures would be better
described with different values
(differences in the electronic specific heat
coefficient $\gamma$ and in the Pu-Ga distances
in the low-$T$ and high-$T$ crystal structure
suggest different degrees of $5f$ delocalization,\cite{boulet05} 
implying incompatible values of $U$).

\begin{table}
\begin{center}
\begin{tabular}{c | c  c c c | c  }
structure & \multispan4{\hfil $U$ (eV)\hfil} & SOC \cr
   & 0 & 1 &  2 & 3 & ($U=0$)\cr
\hline
\multispan6{\hfil Relative energies (meV/PuGa$_3$)\hfil} \cr
\hline
low-$T$    &  0 & 0 &  0  & 0    &  0\cr
``$\alpha$-La"  &  -49  & -9 & 6 & 109   &  -26 \cr
high-$T$ &  -72 & -53 & -26 & 120            &  -47  \cr
AuCu$_3$  & -117  & -67 & 2 & 168    &  -62  \cr
\hline
 \multispan6{\hfil Volumes (\AA$^3$/PuGa$_3$)\hfil} \cr
\hline
 low-$T$   &  77.37 & 78.09 & 80.00 & 81.28  & 77.99 \cr
 ``$\alpha$-La" &  77.59 & 78.58 & 79.78 & 80.84         & 77.52 \cr
high-$T$ &  77.50 & 78.22 & 79.68 & 80.90                       & 77.50 \cr
AuCu$_3$ &  77.72 & 77.72 & 79.88 & 80.77               & 77.67 \cr
\hline
\multispan6{\hfil $c/a$ ratio \hfil} \cr
\hline
 low-$T$    & 0.38 & 0.38 & 0.37 & 0.36 & 0.37  \cr
 ``$\alpha$-La" & 0.40 & 0.39 & 0.39 & 0.39         & 0.39  \cr
high-$T$ & 0.39 & 0.39 & 0.39 & 0.38                       & 0.39 \cr
AuCu$_3$ & 0.41 & 0.41 & 0.41 & 0.41               & 0.41  \cr
\end{tabular}
\caption[]{
Relative energies, volumes, and $c/a$ ratios
for the structures calculated with DFT in the GGA
with different values for $U$ (without spin-orbit coupling)
and for $U=0$ eV with spin-orbit coupling (SOC).
All results are for the AFM state.
The measured values
for the volume are 78.12 and 77.70 \AA$^3$
and for the $c/a$ ratio are 0.358 and 0.378
for the low-$T$ and high-$T$ structures, respectively.\cite{boulet05}
}
\label{tab:opt}
\end{center}
\end{table}

\begin{figure} 
\includegraphics[width=8.5cm]{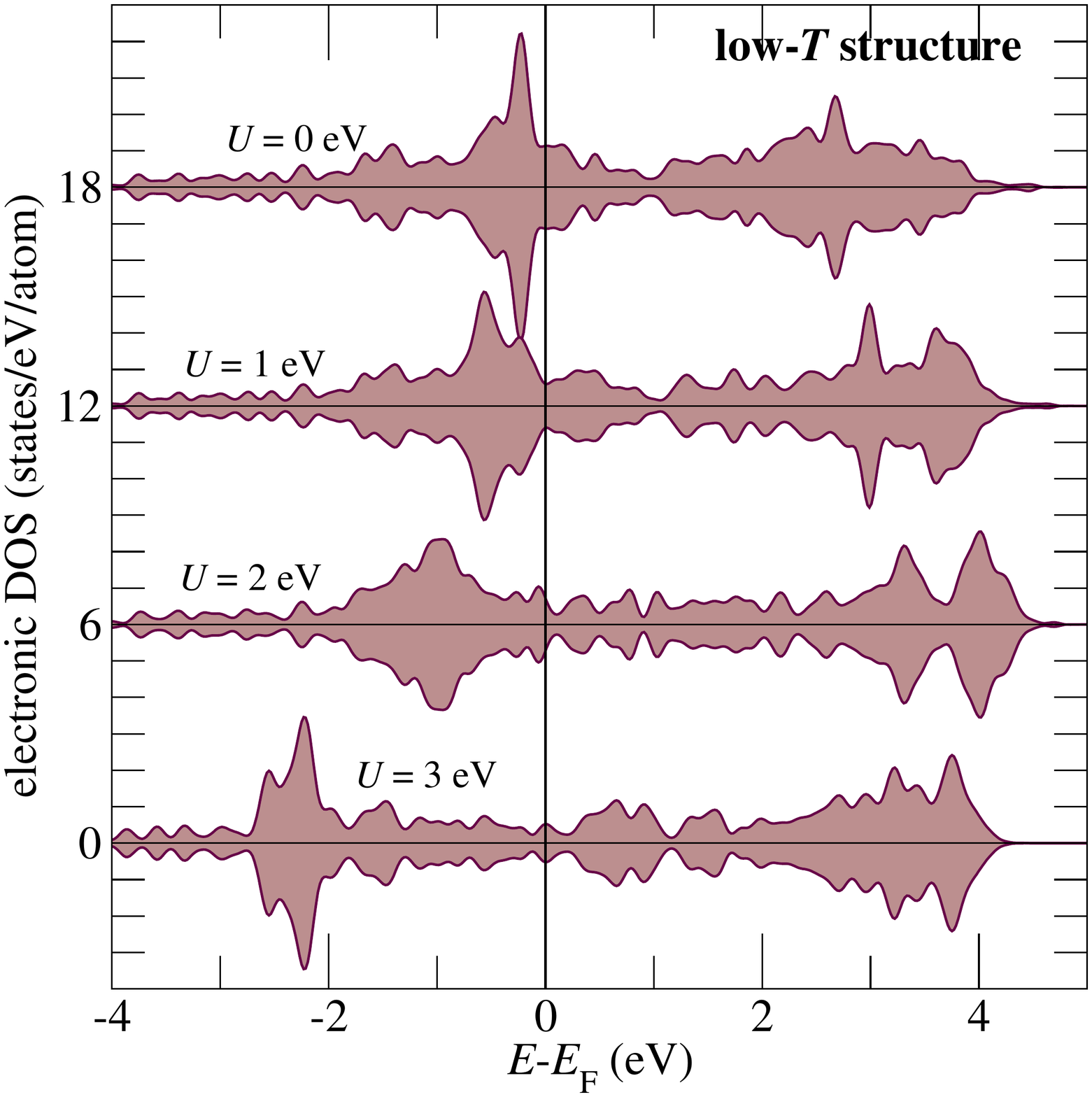}
\caption{\label{fig:eDOSABaf1}
(Color online)
Calculated electronic DOS  
with varying Hubbard $U$ for
the low-$T$ crystal structure with AFM magnetic structure
at the experimental volume.
The DOS for the two spin orientations appear as positive and negative, respectively.
}
\end{figure}

Figure \ref{fig:eDOSABaf1} plots the electronic DOS for the
low-$T$ crystal structure in the AFM state calculated with the GGA+U method.
As $U$ increases, the dominant peaks,
DFT's rendering of the upper and lower Hubbard bands,\cite{hubbard1963}
increasingly separate.
This separation pushes the occupied states down from $E_\text{F}$
more than it pushes the unoccupied states up.
The symmetry between up and down spin remains intact,
and, based on site-projected DOS (not shown here),
the equivalence among sites with the same spin remains.

\begin{figure} 
\includegraphics[width=8.5cm]{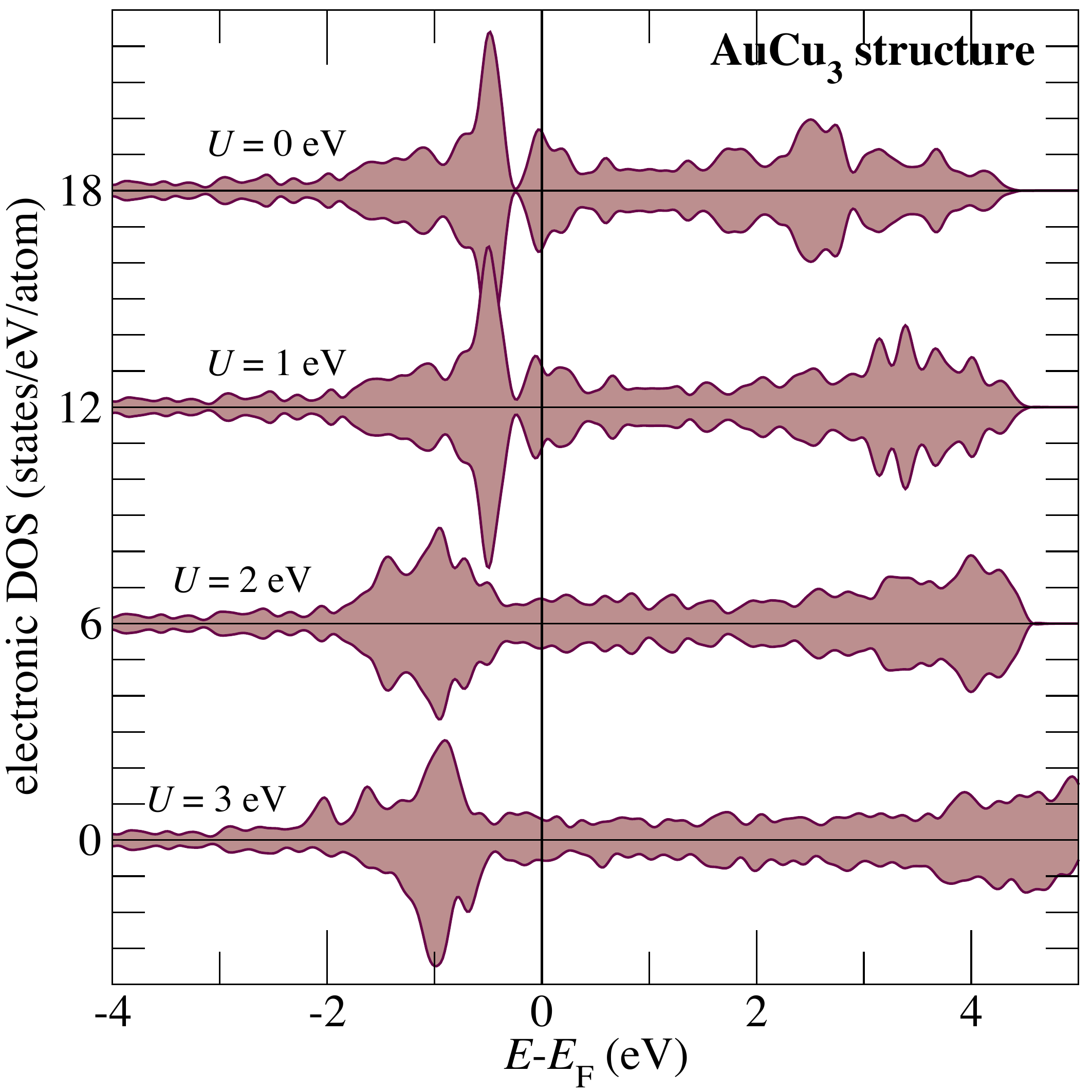}
\caption{\label{fig:eDOSABCaf1}
(Color online)
Calculated electronic DOS  
with varying Hubbard $U$ for
the cubic crystal structure with AFM magnetic structure
at the experimental volume.
The DOS for the two spin orientations appear as positive and negative, respectively.
For U=$2$ eV and above the symmetry between Pu sites is broken and variations of
up to 0.6\% appear in the site-projected charge
and of up to 5\% appear in the site-projected magnetic moments.
}
\end{figure}

Figure \ref{fig:eDOSABCaf1} plots the electronic DOS for the
AuCu$_3$ crystal structure in the AFM state calculated with the GGA+U method.
Again the increasing $U$ drives the dominant peaks apart, but
for this crystal structure the separation occurs mainly by pushing up the
unoccupied states.
The occupied states change little as $U$ increases from 0 eV to 1 eV.
Increasing $U$ from 1 eV to 2 eV pushes the occupied states down.
Setting $U=3$ eV breaks the symmetries of up and down spins
as well as the equivalence among sites with the same spin.

Comparison of Figs. \ref{fig:eDOSABaf1} and \ref{fig:eDOSABCaf1}
for each value of $U$ correlates well with the energy differences
in Table \ref{tab:opt}.
For $U=1$ eV
the AuCu$_3$ crystal structure retains the peak around 0.5 eV
below $E_\text{F}$ while
the low-$T$ crystal structure has its main peak shifted lower than for $U=0$.
For $U=2$ eV
both crystal structures have shifted (and broadened) the peak to
around 1 eV below $E_\text{F}$.
For $U=3$ eV
the overall DOS changes somewhat for the AuCu$_3$ crystal structure
while for the low-$T$ crystal structure a dramatic shift downward occurs.

\section{discussion and conclusions}

Systematically studying the crystal and magnetic structures of PuGa$_3$
reveals how they affect the electronic structure and how the three types
of structure are entangled.
The key to understanding the entanglement lies in
the position of the $5f$ electron states relative to the
Fermi level $E_\text{F}$ in the electronic DOS.
The position relative to $E_\text{F}$ is determined by both the symmetry of
the crystal structure and the imposed magnetic structure.
How far the $5f$ peak sits below $E_\text{F}$ dovetails with how
favorable the system in question is in terms of calculated total energy.

Magnetic structure affects the energy more strongly than crystal structure.
With or without spin-orbit coupling,
the calculations favor the AFM state
over the FM state for all crystal structures.
Spin density waves with wave lengths between those of the FM state (infinity)
and of the AFM state (twice the spacing between Pu planes)
give total energies between the two limiting values.
Pu sites neighboring a junction between up and down spins have their
$5f$ electron states farther below $E_\text{F}$ than other Pu sites.
Each such junction gives the sites sandwiching the junction less hybridization
of $5f$ states with neighbors on the other side of the junction.
In the limiting case of AFM, every site has the least hybridization
because Pu sites in neighboring planes have opposite spin.

Calculations using standard GGA result in the wrong crystal structure
(AuCu$_3$) having the lowest energy in the favored AFM state.
The $5f$ electron states 
in the cubic structure sit farther below $E_\text{F}$ than they do in
the experimentally observed Ni$_3$Sn crystal structure,
because the inversion symmetry at sites in the cubic structure requires
less hybridization between the Pu $5f$ states and other states.

Adding a Hubbard U to treat the strong $5f$ electron correlation results
in the correct crystal structure having the lowest energy.
The $U$ raises and lowers the potential acting on the 
unoccupied and occupied $5f$ states, respectively, but the
effect of $U$ on the positions of the $5f$ states relative to $E_\text{F}$
depends on how they are hybridized.\cite{dudarev97}
Increasing the value of $U$ proves more effective at lowering the
energy of the $5f$ electron peak for the Ni$_3$Sn crystal structure,
making it most favored for $U=3$ eV.

Allowing localized magnetic moments to simulate correlation
effects fails for PuGa$_3$.
The strong preference for the cubic crystal structure over the hexagonal
crystal structure suggests the failure stems not from the actual presence
of a magnetic structure (observed in experiment), but from the
symmetry at Pu sites in the hexagonal crystal structure being much
lower than in the cubic crystal structure.
The use of allowing localized magnetic moments to simulate correlation
does so by permitting the $5f$ electrons on the same Pu site to occupy more orbitals that
differ spatially.
The inversion symmetry present in the cubic crystal structure makes the
localized magnetic moments most effective at simulating correlation
effects.
In the hexagonal crystal structure the lower symmetry prevents an
adequate decoupling of $f$ states from hybridization and their
energy cannot be lowered sufficiently to make the crystal structure
most favorable.

These results suggest a explanation for the effectiveness of
using magnetism to
approximate correlation effects in $\delta$-Pu.
The crystal structure of $\delta$-Pu is face-centered cubic, and all sites
exhibit the inversion symmetry shown here to be important in the
closely-related AuCu$_3$ structure.
Given the similarities, the preference for an AFM state in $\delta$-Pu
does not surprise.
Nor does the additional effectiveness of modeling correlation effects
with disordered local moments astonish, since such a magnetic ``structure''
reduces also the in-plane
hybridization between $f$ electrons on neighboring sites.

Analogous to the relation between
$\delta$-Pu and PuGa$_3$ in the AuCu$_3$ crystal structure,
$\alpha$-Pu relates to PuGa$_3$ in the Ni$_3$Sn crystal structure.
The crystal structures of
both $\alpha$-Pu and the low-$T$ phase of PuGa$_3$ are the most stable
and both have an AB stacking pattern.
The $\alpha$-Pu crystal structure stacks distorted close-packed Pu planes;
replacing three of four Pu atoms with Ga removes the distortion to
restore the symmetry in the close packed planes of PuGa$_3$,
which could relate to the stabilization of $\delta$-Pu to low temperatures by
adding a small amount of Ga.\cite{pluto}
The electronic specific heat coefficient $\gamma$
differs dramatically between the low-$T$ phase of PuGa$_3$,
where $\gamma=220$ mJ/mol K$^2$, and $\alpha$-Pu,
where $\gamma=17$ mJ/mol K$^2$ was measured.\cite{lashley03}
Correspondingly, $\alpha$-Pu can be well described by standard DFT
methods,\cite{soderlind97}
while the work presented here shows that the low-$T$ phase of PuGa$_3$
requires special attention be paid to the strong $5f$ electron correlation,
and allowing spin polarization does not suffice to describe the
effects of the strong correlation.

\begin{acknowledgments}

This research was supported by the Los Alamos National Laboratory,
under the auspices of the National Nuclear Security Agency,
by the U.S. Department of Energy under Grant No.
LDRD-DR 20120024 (``Pu-242: A National Resource for the
Fundamental Understanding of the 5f Electrons of Pu'').
Many thanks go to
in particular 
Eric Chisolm,
Anders Niklasson,
and John Wills
as well as
Eric Bauer,
John Joyce,
and 
Paul Tobash,
for helpful and encouraging discussions.
The author expresses a deep gratitude to Neil Henson
for assistance with the {\sc andulu} computational facility.
Last, but not least, fond thanks go to Lucia Li\^en and Anna Lan for
spurring alternative approaches to understanding.

\end{acknowledgments}


\begin{thebibliography}{26}%
\makeatletter
\providecommand \@ifxundefined [1]{%
 \@ifx{#1\undefined}
}%
\providecommand \@ifnum [1]{%
 \ifnum #1\expandafter \@firstoftwo
 \else \expandafter \@secondoftwo
 \fi
}%
\providecommand \@ifx [1]{%
 \ifx #1\expandafter \@firstoftwo
 \else \expandafter \@secondoftwo
 \fi
}%
\providecommand \natexlab [1]{#1}%
\providecommand \enquote  [1]{``#1''}%
\providecommand \bibnamefont  [1]{#1}%
\providecommand \bibfnamefont [1]{#1}%
\providecommand \citenamefont [1]{#1}%
\providecommand \href@noop [0]{\@secondoftwo}%
\providecommand \href [0]{\begingroup \@sanitize@url \@href}%
\providecommand \@href[1]{\@@startlink{#1}\@@href}%
\providecommand \@@href[1]{\endgroup#1\@@endlink}%
\providecommand \@sanitize@url [0]{\catcode `\\12\catcode `\$12\catcode
  `\&12\catcode `\#12\catcode `\^12\catcode `\_12\catcode `\%12\relax}%
\providecommand \@@startlink[1]{}%
\providecommand \@@endlink[0]{}%
\providecommand \url  [0]{\begingroup\@sanitize@url \@url }%
\providecommand \@url [1]{\endgroup\@href {#1}{\urlprefix }}%
\providecommand \urlprefix  [0]{URL }%
\providecommand \Eprint [0]{\href }%
\providecommand \doibase [0]{http://dx.doi.org/}%
\providecommand \selectlanguage [0]{\@gobble}%
\providecommand \bibinfo  [0]{\@secondoftwo}%
\providecommand \bibfield  [0]{\@secondoftwo}%
\providecommand \translation [1]{[#1]}%
\providecommand \BibitemOpen [0]{}%
\providecommand \bibitemStop [0]{}%
\providecommand \bibitemNoStop [0]{.\EOS\space}%
\providecommand \EOS [0]{\spacefactor3000\relax}%
\providecommand \BibitemShut  [1]{\csname bibitem#1\endcsname}%
\let\auto@bib@innerbib\@empty
\bibitem [{\citenamefont {Ellinger}\ \emph {et~al.}(1964)\citenamefont
  {Ellinger}, \citenamefont {Land},\ and\ \citenamefont
  {Struebing}}]{ellinger64}%
  \BibitemOpen
  \bibfield  {author} {\bibinfo {author} {\bibfnamefont {F.~H.}\ \bibnamefont
  {Ellinger}}, \bibinfo {author} {\bibfnamefont {C.~C.}\ \bibnamefont {Land}},
  \ and\ \bibinfo {author} {\bibfnamefont {V.~O.}\ \bibnamefont {Struebing}},\
  }\href@noop {} {\bibfield  {journal} {\bibinfo  {journal} {J. Nucl. Mater.}\
  }\textbf {\bibinfo {volume} {12}},\ \bibinfo {pages} {226} (\bibinfo {year}
  {1964})}\BibitemShut {NoStop}%
\bibitem [{\citenamefont {Larson}\ \emph {et~al.}(1965)\citenamefont {Larson},
  \citenamefont {Cromer},\ and\ \citenamefont {Roof}}]{larson1965}%
  \BibitemOpen
  \bibfield  {author} {\bibinfo {author} {\bibfnamefont {A.~C.}\ \bibnamefont
  {Larson}}, \bibinfo {author} {\bibfnamefont {D.~T.}\ \bibnamefont {Cromer}},
  \ and\ \bibinfo {author} {\bibfnamefont {R.~B.}\ \bibnamefont {Roof},
  \bibfnamefont {Jnr}},\ }\href@noop {} {\bibfield  {journal} {\bibinfo
  {journal} {Acta Crystallographica}\ }\textbf {\bibinfo {volume} {18}},\
  \bibinfo {pages} {294} (\bibinfo {year} {1965})}\BibitemShut {NoStop}%
\bibitem [{\citenamefont {Boulet}\ \emph {et~al.}(2005)\citenamefont {Boulet},
  \citenamefont {Colineau}, \citenamefont {Wastin}, \citenamefont {Javorsk\'y},
  \citenamefont {Griveau}, \citenamefont {Rebizant}, \citenamefont {Stewart},\
  and\ \citenamefont {Bauer}}]{boulet05}%
  \BibitemOpen
  \bibfield  {author} {\bibinfo {author} {\bibfnamefont {P.}~\bibnamefont
  {Boulet}}, \bibinfo {author} {\bibfnamefont {E.}~\bibnamefont {Colineau}},
  \bibinfo {author} {\bibfnamefont {F.}~\bibnamefont {Wastin}}, \bibinfo
  {author} {\bibfnamefont {P.}~\bibnamefont {Javorsk\'y}}, \bibinfo {author}
  {\bibfnamefont {J.~C.}\ \bibnamefont {Griveau}}, \bibinfo {author}
  {\bibfnamefont {J.}~\bibnamefont {Rebizant}}, \bibinfo {author}
  {\bibfnamefont {G.~R.}\ \bibnamefont {Stewart}}, \ and\ \bibinfo {author}
  {\bibfnamefont {E.~D.}\ \bibnamefont {Bauer}},\ }\href@noop {} {\bibfield
  {journal} {\bibinfo  {journal} {Phys. Rev. B}\ }\textbf {\bibinfo {volume}
  {72}},\ \bibinfo {pages} {064438} (\bibinfo {year} {2005})}\BibitemShut
  {NoStop}%
\bibitem [{\citenamefont {Javorsk\'y}\ \emph {et~al.}(2007)\citenamefont
  {Javorsk\'y}, \citenamefont {Colineau}, \citenamefont {Wastin}, \citenamefont
  {Jutier}, \citenamefont {Griveau}, \citenamefont {Boulet}, \citenamefont
  {Jardin},\ and\ \citenamefont {Rebizant}}]{javorsky2007}%
  \BibitemOpen
  \bibfield  {author} {\bibinfo {author} {\bibfnamefont {P.}~\bibnamefont
  {Javorsk\'y}}, \bibinfo {author} {\bibfnamefont {E.}~\bibnamefont
  {Colineau}}, \bibinfo {author} {\bibfnamefont {F.}~\bibnamefont {Wastin}},
  \bibinfo {author} {\bibfnamefont {F.}~\bibnamefont {Jutier}}, \bibinfo
  {author} {\bibfnamefont {J.-C.}\ \bibnamefont {Griveau}}, \bibinfo {author}
  {\bibfnamefont {P.}~\bibnamefont {Boulet}}, \bibinfo {author} {\bibfnamefont
  {R.}~\bibnamefont {Jardin}}, \ and\ \bibinfo {author} {\bibfnamefont
  {J.}~\bibnamefont {Rebizant}},\ }\href {\doibase 10.1103/PhysRevB.75.184501}
  {\bibfield  {journal} {\bibinfo  {journal} {Phys. Rev. B}\ }\textbf {\bibinfo
  {volume} {75}},\ \bibinfo {pages} {184501} (\bibinfo {year}
  {2007})}\BibitemShut {NoStop}%
\bibitem [{\citenamefont {Opahle}\ \emph {et~al.}(2004)\citenamefont {Opahle},
  \citenamefont {Elgazzar}, \citenamefont {Koepernik},\ and\ \citenamefont
  {Oppeneer}}]{opahle2004}%
  \BibitemOpen
  \bibfield  {author} {\bibinfo {author} {\bibfnamefont {I.}~\bibnamefont
  {Opahle}}, \bibinfo {author} {\bibfnamefont {S.}~\bibnamefont {Elgazzar}},
  \bibinfo {author} {\bibfnamefont {K.}~\bibnamefont {Koepernik}}, \ and\
  \bibinfo {author} {\bibfnamefont {P.~M.}\ \bibnamefont {Oppeneer}},\ }\href
  {\doibase 10.1103/PhysRevB.70.104504} {\bibfield  {journal} {\bibinfo
  {journal} {Phys. Rev. B}\ }\textbf {\bibinfo {volume} {70}},\ \bibinfo
  {pages} {104504} (\bibinfo {year} {2004})}\BibitemShut {NoStop}%
\bibitem [{\citenamefont {Lashley}\ \emph {et~al.}(2005)\citenamefont
  {Lashley}, \citenamefont {Lawson}, \citenamefont {McQueeney},\ and\
  \citenamefont {Lander}}]{lashley05}%
  \BibitemOpen
  \bibfield  {author} {\bibinfo {author} {\bibfnamefont {J.~C.}\ \bibnamefont
  {Lashley}}, \bibinfo {author} {\bibfnamefont {A.}~\bibnamefont {Lawson}},
  \bibinfo {author} {\bibfnamefont {R.~J.}\ \bibnamefont {McQueeney}}, \ and\
  \bibinfo {author} {\bibfnamefont {G.~H.}\ \bibnamefont {Lander}},\
  }\href@noop {} {\bibfield  {journal} {\bibinfo  {journal} {Phys. Rev. B}\
  }\textbf {\bibinfo {volume} {72}},\ \bibinfo {pages} {054416} (\bibinfo
  {year} {2005})}\BibitemShut {NoStop}%
\bibitem [{\citenamefont {Postnikov}\ and\ \citenamefont
  {Antropov}(2000)}]{postinov00}%
  \BibitemOpen
  \bibfield  {author} {\bibinfo {author} {\bibfnamefont {A.~V.}\ \bibnamefont
  {Postnikov}}\ and\ \bibinfo {author} {\bibfnamefont {V.~P.}\ \bibnamefont
  {Antropov}},\ }\href@noop {} {\bibfield  {journal} {\bibinfo  {journal}
  {Comp. Mat. Science}\ }\textbf {\bibinfo {volume} {17}},\ \bibinfo {pages}
  {438} (\bibinfo {year} {2000})}\BibitemShut {NoStop}%
\bibitem [{\citenamefont {Wang}\ and\ \citenamefont {Sun}(2000)}]{wang00}%
  \BibitemOpen
  \bibfield  {author} {\bibinfo {author} {\bibfnamefont {Y.}~\bibnamefont
  {Wang}}\ and\ \bibinfo {author} {\bibfnamefont {Y.}~\bibnamefont {Sun}},\
  }\href@noop {} {\bibfield  {journal} {\bibinfo  {journal} {J. Phys.: Condens.
  Matter}\ }\textbf {\bibinfo {volume} {12}},\ \bibinfo {pages} {L311}
  (\bibinfo {year} {2000})}\BibitemShut {NoStop}%
\bibitem [{\citenamefont {Soderlind}(2001)}]{soderlind01}%
  \BibitemOpen
  \bibfield  {author} {\bibinfo {author} {\bibfnamefont {P.}~\bibnamefont
  {Soderlind}},\ }\href@noop {} {\bibfield  {journal} {\bibinfo  {journal}
  {Eur. Phys. Lett.}\ }\textbf {\bibinfo {volume} {55}},\ \bibinfo {pages}
  {525} (\bibinfo {year} {2001})}\BibitemShut {NoStop}%
\bibitem [{\citenamefont {Niklasson}\ \emph {et~al.}(2003)\citenamefont
  {Niklasson}, \citenamefont {Wills}, \citenamefont {Katsnelson}, \citenamefont
  {Abrikosov}, \citenamefont {Eriksson},\ and\ \citenamefont
  {Johansson}}]{niklasson03}%
  \BibitemOpen
  \bibfield  {author} {\bibinfo {author} {\bibfnamefont {A.~M.~N.}\
  \bibnamefont {Niklasson}}, \bibinfo {author} {\bibfnamefont {J.~M.}\
  \bibnamefont {Wills}}, \bibinfo {author} {\bibfnamefont {M.~I.}\ \bibnamefont
  {Katsnelson}}, \bibinfo {author} {\bibfnamefont {I.~A.}\ \bibnamefont
  {Abrikosov}}, \bibinfo {author} {\bibfnamefont {O.}~\bibnamefont {Eriksson}},
  \ and\ \bibinfo {author} {\bibfnamefont {B.}~\bibnamefont {Johansson}},\
  }\href@noop {} {\bibfield  {journal} {\bibinfo  {journal} {Phys.\ Rev.\ B}\
  }\textbf {\bibinfo {volume} {67}},\ \bibinfo {pages} {235105} (\bibinfo
  {year} {2003})}\BibitemShut {NoStop}%
\bibitem [{\citenamefont {Soderlind}\ \emph {et~al.}(2004)\citenamefont
  {Soderlind}, \citenamefont {Landa}, \citenamefont {Sadigh}, \citenamefont
  {Vitos},\ and\ \citenamefont {Ruban}}]{soderlind04a}%
  \BibitemOpen
  \bibfield  {author} {\bibinfo {author} {\bibfnamefont {P.}~\bibnamefont
  {Soderlind}}, \bibinfo {author} {\bibfnamefont {A.}~\bibnamefont {Landa}},
  \bibinfo {author} {\bibfnamefont {B.}~\bibnamefont {Sadigh}}, \bibinfo
  {author} {\bibfnamefont {L.}~\bibnamefont {Vitos}}, \ and\ \bibinfo {author}
  {\bibfnamefont {A.}~\bibnamefont {Ruban}},\ }\href@noop {} {\bibfield
  {journal} {\bibinfo  {journal} {Phys. Rev. B}\ }\textbf {\bibinfo {volume}
  {70}},\ \bibinfo {pages} {144103} (\bibinfo {year} {2004})}\BibitemShut
  {NoStop}%
\bibitem [{\citenamefont {Crocker}(1971)}]{crocker1971}%
  \BibitemOpen
  \bibfield  {author} {\bibinfo {author} {\bibfnamefont {A.}~\bibnamefont
  {Crocker}},\ }\href@noop {} {\bibfield  {journal} {\bibinfo  {journal}
  {Journal of Nuclear Materials}\ }\textbf {\bibinfo {volume} {41}},\ \bibinfo
  {pages} {167 } (\bibinfo {year} {1971})}\BibitemShut {NoStop}%
\bibitem [{\citenamefont {Kresse}\ and\ \citenamefont
  {Furthmuller}(1996)}]{kresse96}%
  \BibitemOpen
  \bibfield  {author} {\bibinfo {author} {\bibfnamefont {G.}~\bibnamefont
  {Kresse}}\ and\ \bibinfo {author} {\bibfnamefont {J.}~\bibnamefont
  {Furthmuller}},\ }\href@noop {} {\bibfield  {journal} {\bibinfo  {journal}
  {Phys. Rev. B}\ }\textbf {\bibinfo {volume} {54}},\ \bibinfo {pages} {11169}
  (\bibinfo {year} {1996})}\BibitemShut {NoStop}%
\bibitem [{\citenamefont {Kresse}\ and\ \citenamefont
  {Joubert}(1999)}]{kresse99}%
  \BibitemOpen
  \bibfield  {author} {\bibinfo {author} {\bibfnamefont {G.}~\bibnamefont
  {Kresse}}\ and\ \bibinfo {author} {\bibfnamefont {D.}~\bibnamefont
  {Joubert}},\ }\href@noop {} {\bibfield  {journal} {\bibinfo  {journal} {Phys.
  Rev. B}\ }\textbf {\bibinfo {volume} {59}},\ \bibinfo {pages} {1758}
  (\bibinfo {year} {1999})}\BibitemShut {NoStop}%
\bibitem [{\citenamefont {Perdew}\ \emph {et~al.}(1996)\citenamefont {Perdew},
  \citenamefont {Burke},\ and\ \citenamefont {Ernzerhof}}]{PBE96}%
  \BibitemOpen
  \bibfield  {author} {\bibinfo {author} {\bibfnamefont {J.~P.}\ \bibnamefont
  {Perdew}}, \bibinfo {author} {\bibfnamefont {K.}~\bibnamefont {Burke}}, \
  and\ \bibinfo {author} {\bibfnamefont {M.}~\bibnamefont {Ernzerhof}},\
  }\href@noop {} {\bibfield  {journal} {\bibinfo  {journal} {Phys.\ Rev.\
  Lett.}\ }\textbf {\bibinfo {volume} {77}},\ \bibinfo {pages} {3865} (\bibinfo
  {year} {1996})}\BibitemShut {NoStop}%
\bibitem [{\citenamefont {Bl{\"o}chl}(1994)}]{blochl94a}%
  \BibitemOpen
  \bibfield  {author} {\bibinfo {author} {\bibfnamefont {P.~E.}\ \bibnamefont
  {Bl{\"o}chl}},\ }\href@noop {} {\bibfield  {journal} {\bibinfo  {journal}
  {Phys. Rev. B}\ }\textbf {\bibinfo {volume} {50}},\ \bibinfo {pages} {17953}
  (\bibinfo {year} {1994})}\BibitemShut {NoStop}%
\bibitem [{\citenamefont {Dudarev}\ \emph {et~al.}(1998)\citenamefont
  {Dudarev}, \citenamefont {Botton}, \citenamefont {Savrasov}, \citenamefont
  {Humphreys},\ and\ \citenamefont {Sutton}}]{dudarev98}%
  \BibitemOpen
  \bibfield  {author} {\bibinfo {author} {\bibfnamefont {S.~L.}\ \bibnamefont
  {Dudarev}}, \bibinfo {author} {\bibfnamefont {G.~A.}\ \bibnamefont {Botton}},
  \bibinfo {author} {\bibfnamefont {S.~Y.}\ \bibnamefont {Savrasov}}, \bibinfo
  {author} {\bibfnamefont {C.~J.}\ \bibnamefont {Humphreys}}, \ and\ \bibinfo
  {author} {\bibfnamefont {A.~P.}\ \bibnamefont {Sutton}},\ }\href@noop {}
  {\bibfield  {journal} {\bibinfo  {journal} {Phys. Rev. B}\ }\textbf {\bibinfo
  {volume} {57}},\ \bibinfo {pages} {1505} (\bibinfo {year}
  {1998})}\BibitemShut {NoStop}%
\bibitem [{\citenamefont {Hobbs}\ \emph {et~al.}(2000)\citenamefont {Hobbs},
  \citenamefont {Kresse},\ and\ \citenamefont {Hafner}}]{hobbs00}%
  \BibitemOpen
  \bibfield  {author} {\bibinfo {author} {\bibfnamefont {D.}~\bibnamefont
  {Hobbs}}, \bibinfo {author} {\bibfnamefont {G.}~\bibnamefont {Kresse}}, \
  and\ \bibinfo {author} {\bibfnamefont {J.}~\bibnamefont {Hafner}},\
  }\href@noop {} {\bibfield  {journal} {\bibinfo  {journal} {Phys.\ Rev.\ B}\
  }\textbf {\bibinfo {volume} {62}},\ \bibinfo {pages} {11556} (\bibinfo {year}
  {2000})}\BibitemShut {NoStop}%
\bibitem [{\citenamefont {Marsman}\ and\ \citenamefont
  {Hafner}(2002)}]{marsman02}%
  \BibitemOpen
  \bibfield  {author} {\bibinfo {author} {\bibfnamefont {M.}~\bibnamefont
  {Marsman}}\ and\ \bibinfo {author} {\bibfnamefont {J.}~\bibnamefont
  {Hafner}},\ }\href@noop {} {\bibfield  {journal} {\bibinfo  {journal} {Phys.
  Rev. B}\ }\textbf {\bibinfo {volume} {66}},\ \bibinfo {pages} {224409}
  (\bibinfo {year} {2002})}\BibitemShut {NoStop}%
\bibitem [{\citenamefont {Kleinman}(1980)}]{PhysRevB.21.2630}%
  \BibitemOpen
  \bibfield  {author} {\bibinfo {author} {\bibfnamefont {L.}~\bibnamefont
  {Kleinman}},\ }\href {\doibase 10.1103/PhysRevB.21.2630} {\bibfield
  {journal} {\bibinfo  {journal} {Phys. Rev. B}\ }\textbf {\bibinfo {volume}
  {21}},\ \bibinfo {pages} {2630} (\bibinfo {year} {1980})}\BibitemShut
  {NoStop}%
\bibitem [{\citenamefont {MacDonald}\ \emph {et~al.}(1980)\citenamefont
  {MacDonald}, \citenamefont {Picket},\ and\ \citenamefont
  {Koelling}}]{macdonald80}%
  \BibitemOpen
  \bibfield  {author} {\bibinfo {author} {\bibfnamefont {A.~H.}\ \bibnamefont
  {MacDonald}}, \bibinfo {author} {\bibfnamefont {W.~E.}\ \bibnamefont
  {Picket}}, \ and\ \bibinfo {author} {\bibfnamefont {D.~D.}\ \bibnamefont
  {Koelling}},\ }\href {http://stacks.iop.org/0022-3719/13/i=14/a=009}
  {\bibfield  {journal} {\bibinfo  {journal} {Journal of Physics C: Solid State
  Physics}\ }\textbf {\bibinfo {volume} {13}},\ \bibinfo {pages} {2675}
  (\bibinfo {year} {1980})}\BibitemShut {NoStop}%
\bibitem [{\citenamefont {Hubbard}(1963)}]{hubbard1963}%
  \BibitemOpen
  \bibfield  {author} {\bibinfo {author} {\bibfnamefont {J.}~\bibnamefont
  {Hubbard}},\ }\href@noop {} {\bibfield  {journal} {\bibinfo  {journal}
  {Proceedings of the Royal Society of London. Series A, Mathematical and
  Physical Sciences}\ }\textbf {\bibinfo {volume} {276}},\ \bibinfo {pages}
  {238} (\bibinfo {year} {1963})}\BibitemShut {NoStop}%
\bibitem [{\citenamefont {Dudarev}\ \emph {et~al.}(1997)\citenamefont
  {Dudarev}, \citenamefont {Manh},\ and\ \citenamefont {Sutton}}]{dudarev97}%
  \BibitemOpen
  \bibfield  {author} {\bibinfo {author} {\bibfnamefont {S.~L.}\ \bibnamefont
  {Dudarev}}, \bibinfo {author} {\bibfnamefont {D.~N.}\ \bibnamefont {Manh}}, \
  and\ \bibinfo {author} {\bibfnamefont {A.~P.}\ \bibnamefont {Sutton}},\
  }\href@noop {} {\bibfield  {journal} {\bibinfo  {journal} {Philosophical
  Magazine Part B}\ }\textbf {\bibinfo {volume} {75}},\ \bibinfo {pages} {613}
  (\bibinfo {year} {1997})}\BibitemShut {NoStop}%
\bibitem [{\citenamefont {Wick}(1967)}]{pluto}%
  \BibitemOpen
  \bibfield  {author} {\bibinfo {author} {\bibfnamefont {O.~J.}\ \bibnamefont
  {Wick}},\ }\href@noop {} {\emph {\bibinfo {title} {Plutonium Handbook, A
  Guide To The Technology}}}\ (\bibinfo  {publisher} {Gordon and Breach},\
  \bibinfo {address} {New York},\ \bibinfo {year} {1967})\BibitemShut {NoStop}%
\bibitem [{\citenamefont {Lashley}\ \emph {et~al.}(2003)\citenamefont
  {Lashley}, \citenamefont {Singleton}, \citenamefont {Migliori}, \citenamefont
  {Betts}, \citenamefont {Fisher}, \citenamefont {Smith},\ and\ \citenamefont
  {McQueeney}}]{lashley03}%
  \BibitemOpen
  \bibfield  {author} {\bibinfo {author} {\bibfnamefont {J.~C.}\ \bibnamefont
  {Lashley}}, \bibinfo {author} {\bibfnamefont {J.}~\bibnamefont {Singleton}},
  \bibinfo {author} {\bibfnamefont {A.}~\bibnamefont {Migliori}}, \bibinfo
  {author} {\bibfnamefont {J.~B.}\ \bibnamefont {Betts}}, \bibinfo {author}
  {\bibfnamefont {R.~A.}\ \bibnamefont {Fisher}}, \bibinfo {author}
  {\bibfnamefont {J.~L.}\ \bibnamefont {Smith}}, \ and\ \bibinfo {author}
  {\bibfnamefont {R.~J.}\ \bibnamefont {McQueeney}},\ }\href@noop {} {\bibfield
   {journal} {\bibinfo  {journal} {Phys. Rev. Lett.}\ }\textbf {\bibinfo
  {volume} {91}},\ \bibinfo {pages} {205901} (\bibinfo {year}
  {2003})}\BibitemShut {NoStop}%
\bibitem [{\citenamefont {Soderlind}\ \emph {et~al.}(1997)\citenamefont
  {Soderlind}, \citenamefont {Wills}, \citenamefont {Johansson},\ and\
  \citenamefont {Eriksson}}]{soderlind97}%
  \BibitemOpen
  \bibfield  {author} {\bibinfo {author} {\bibfnamefont {P.}~\bibnamefont
  {Soderlind}}, \bibinfo {author} {\bibfnamefont {J.~M.}\ \bibnamefont
  {Wills}}, \bibinfo {author} {\bibfnamefont {B.}~\bibnamefont {Johansson}}, \
  and\ \bibinfo {author} {\bibfnamefont {O.}~\bibnamefont {Eriksson}},\
  }\href@noop {} {\bibfield  {journal} {\bibinfo  {journal} {Phys. Rev. B}\
  }\textbf {\bibinfo {volume} {55}},\ \bibinfo {pages} {1997} (\bibinfo {year}
  {1997})}\BibitemShut {NoStop}%
\end{thebibliography}
\end{document}